\documentclass{iau}
\usepackage{graphicx,natbib}
 
\newcommand{\apj}{ApJ}           
\newcommand{\mnras}{MNRAS}       

\newcommand{\aap}{A\&A}

\newcommand{\aj}{AJ}

\newcommand{\arcsec}{\mbox{\ensuremath{^{\prime\prime}}}}
\newcommand{\farcs}{\mbox{\ensuremath{.\!\!^{\prime\prime}}}}

\title{The SAMI Galaxy Survey: first 1000 galaxies}

\author[J.\,T.\ Allen et al.]{J.\,T.\ Allen$^{1,2}$ \and the SAMI Galaxy Survey Team$^3$}

\affiliation{$^1$ Sydney Institute for Astronomy (SIfA), School of Physics, The University of Sydney, NSW 2006, Australia
\\$^2$ ARC Centre of Excellence for All-sky Astrophysics (CAASTRO)
\\email: {\tt j.allen@physics.usyd.edu.au}
\\$^3$ Full list of team members is available at {\tt http://sami-survey.org/members}}

\pubyear{2014}
\volume{309}
\jname{Galaxies in 3D across the Universe}
\editors{B. L. Ziegler, F. Combes, H. Dannerbauer, M. Verdugo, eds.}

\begin{document}

\maketitle

\begin{abstract}
The Sydney--AAO Multi-object Integral field spectrograph (SAMI) Galaxy Survey is an ongoing project to obtain integral field spectroscopic observations of $\sim$3400 galaxies by mid-2016. Including the pilot survey, a total of $\sim$1000 galaxies have been observed to date, making the SAMI Galaxy Survey the largest of its kind in existence. This unique dataset allows a wide range of investigations into different aspects of galaxy evolution.

The first public data from the SAMI Galaxy Survey, consisting of 107 galaxies drawn from the full sample, has now been released. By giving early access to SAMI data for the entire research community, we aim to stimulate research across a broad range of topics in galaxy evolution. As the sample continues to grow, the survey will open up a new and unique parameter space for galaxy evolution studies.

\end{abstract}

\firstsection
\section{Introduction}

Large-scale spectroscopic galaxies surveys, such as the Sloan Digital Sky Survey (SDSS; \citealt{York00}) and Galaxy And Mass Assembly (GAMA; \citealt{Driver09,Driver11}) survey, are well established and powerful tools for investigations of galaxy evolution. However, they are limited by only observing a single aperture in the centre of each galaxy, so are inherently unable to investigate the spatially distributed properties of galaxies.

Integral field spectrographs (IFSs) provide the information missed by traditional fibre-based surveys. The spatially resolved spectroscopic data produced by an IFS can be used to measure diverse quantities such as rotation curves, spatial distributions of star formation, and radial variation in stellar populations. Until recently, technical limitations made IFS observations difficult and time-consuming, limiting sample sizes to a few hundred at most, e.g. 260 in ATLAS-3D \citep{Cappellari11b} and $\sim$600 in CALIFA \citep{Sanchez12}.

A massive step forward is now being taken by instruments that can make IFS observations of multiple objects at once. The Sydney--AAO Multi-object Integral field spectrograph (SAMI) is one of the first of such instruments, with 13 integral field units (IFUs), each with a field of view of 15\arcsec, that can be deployed across a 1-degree patrol field \citep{Croom12}. Each IFU consists of a bundle of 61 optical fibres lightly fused to have a high ($\sim$75\%) filling factor \citep{Bryant14a}. SAMI is installed on the 3.9-m Anglo-Australian Telescope (AAT), feeding the existing AAOmega spectrograph \citep{Sharp06}. For the first time, SAMI allows the rapid generation of large samples of IFS observations.

\section{The SAMI Galaxy Survey}

The SAMI Galaxy Survey is an ongoing project using SAMI to obtain integral field spectroscopic data for $\sim$3400 galaxies, with an expected completion date of mid-2016. At the time of writing, $\sim$1000 galaxies have been observed, including the pilot survey of $\sim$100 galaxies. Early science investigations have included studies of galactic winds \citep{Fogarty12,Ho14}, the kinematic morphology--density relation \citep{Fogarty14}, and star formation in dwarf galaxies (Richards et al., MNRAS accepted); see also the related contributions by J.~Bland-Hawthorn, L.\,M.\,R.~Fogarty, I.-T.~Ho, and N.~Scott in this volume.

\subsection{Target selection}

The target galaxies for the SAMI Galaxy Survey are split into two samples: a field and group sample from the Galaxy And Mass Assembly (GAMA) survey G09, G12 and G15 fields, and a cluster sample from a set of eight galaxy clusters. The inclusion of a dedicated cluster sample extends the range of environmental densities to higher values than a simple mass- or luminosity-selected sample of galaxies would provide. Full details of the target selection are provided in \citet{Bryant14b}.

\begin{figure}
\centering
\includegraphics[width=0.5\columnwidth]{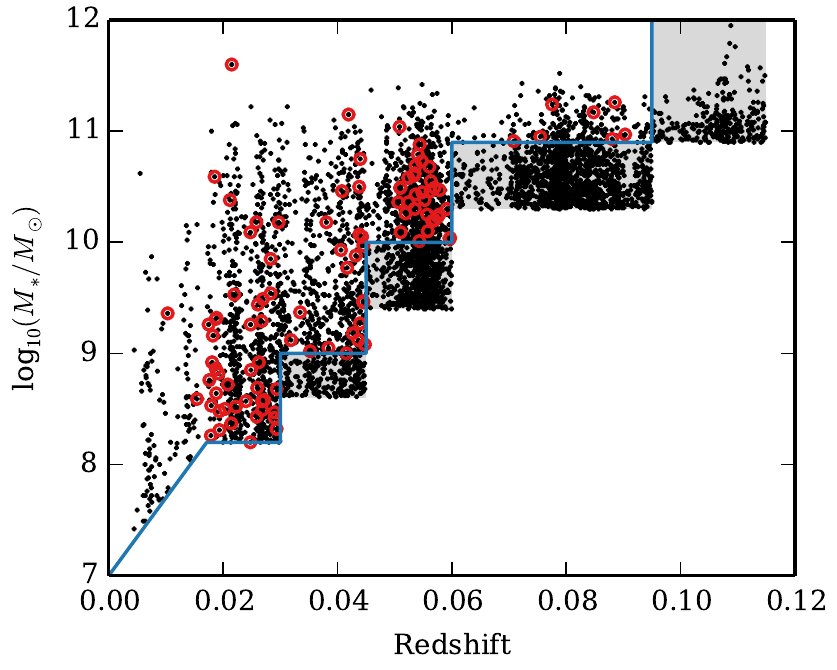} 
\caption{Stellar mass and redshift for all galaxies in the field (GAMA) regions of the SAMI Galaxy Survey (black points) and those in the EDR sample (red circles). The blue boundaries indicate the primary selection criteria, while the shaded regions indicate lower-priority targets. Large-scale structure within the GAMA regions is seen in the overdensities of galaxies at particular redshifts.}\label{fig:target_selection}
\end{figure}

The target selection for the GAMA regions, consisting of a tiered set of volume-limited samples, is illustrated in Fig.~\ref{fig:target_selection}. The figure also marks the galaxies included in the Early Data Release (EDR), discussed in Section~\ref{sec:edr}.

\subsection{Data reduction}

The initial steps in the SAMI data reduction, up to the production of row-stacked spectra (RSS), are carried out using version 5.62 of \textsc{2dfdr}\footnote{\tt http://www.aao.gov.au/science/software/2dfdr}. The dedicated SAMI pipeline \citep{Allen14a} then performs flux calibration using separate observations of spectrophotometric standard stars, and corrects for telluric absorption based on simultaneous observations of a secondary standard star in the galaxy field. The set of $\sim$7 dithered frames in each field are then resampled onto a regular grid and combined to produce a pair of datacubes (representing the blue and red arms of AAOmega) for each galaxy.

\begin{figure}
\centering
\includegraphics[width=\columnwidth]{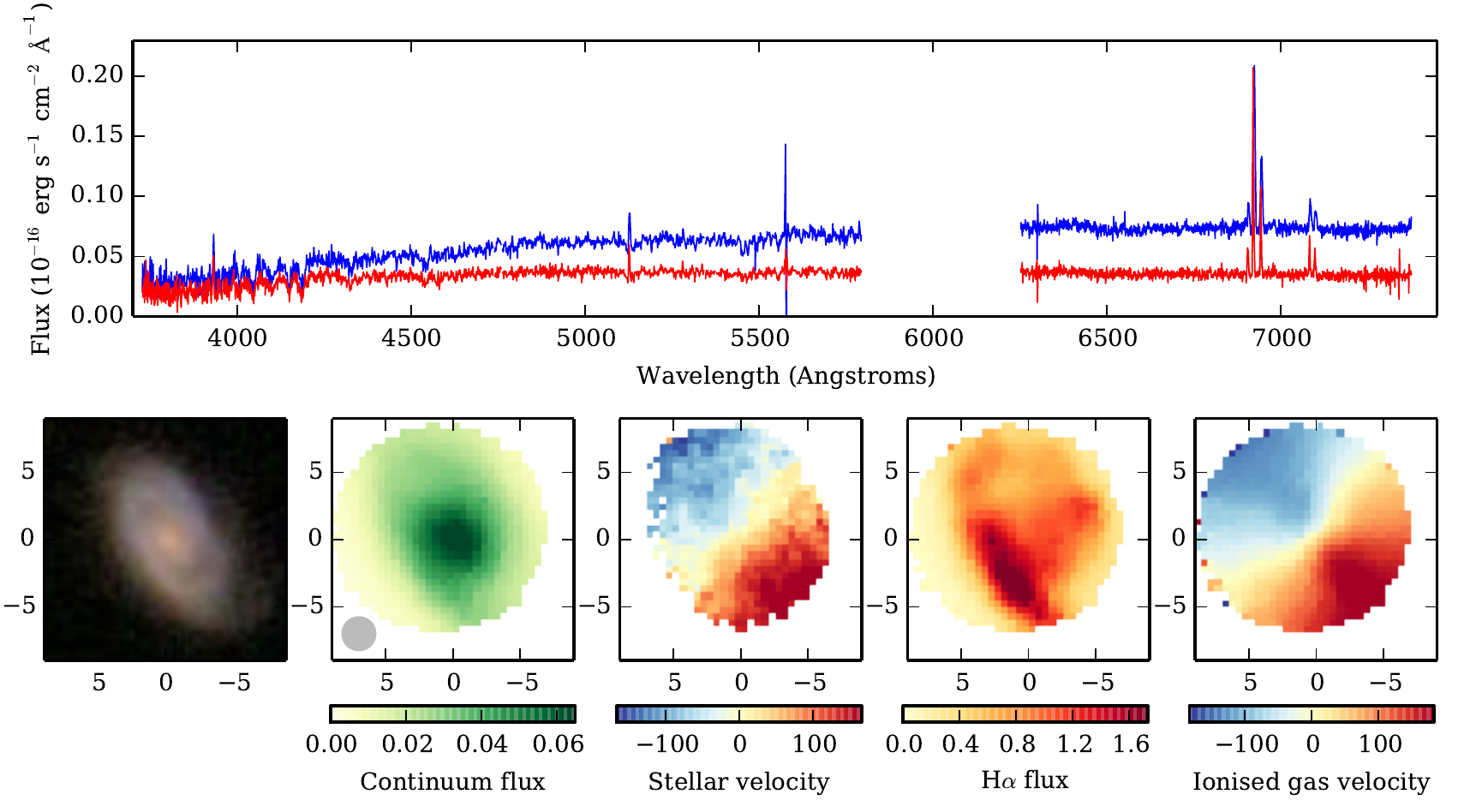} 
\caption{Example SAMI data for the galaxy 511867, with $z=0.05523$ and $M_*=10^{10.68}{\rm M}_\odot$. Upper panel: flux for a central spaxel (blue) and one 3\farcs75 to the North (red). Lower panels, from left to right: SDSS $gri$ image; continuum flux map ($10^{-16}$\,erg\,s$^{-1}$\,cm$^{-2}$\,\AA$^{-1}$); stellar velocity field (km\,s$^{-1}$); H$\alpha$ flux map ($10^{-16}$\,erg\,s$^{-1}$\,cm$^{-2}$); H$\alpha$ velocity field (km\,s$^{-1}$). The two velocity fields are each scaled individually. For the stellar velocity map, only spaxels with per-pixel signal-to-noise ratio $>$5 in the continuum are included. Each panel is 18\arcsec\ square, with North up and East to the left. The grey circle in the second panel shows the FWHM of the PSF.}\label{fig:example_galaxy}
\end{figure}

Fig.~\ref{fig:example_galaxy} illustrates the finished product for a single galaxy, showing the nature and quality of the data obtained. The complete data reduction process is described in \citet{Sharp14}.

\subsection{Data quality}

A comprehensive assessment of the quality of the data produced by SAMI is presented in \citet{Allen14b}; here we provide some key results.

Through the use of dome flat fields we achieve a flat field accuracy of 0.5--1\% in most of the blue arm and 0.3--0.4\% in the red arm. We use a combination of CuAr arc lamps and sky lines to achieve a wavelength calibration accuracy of $\simeq$0.1 pixels or better. The continuum sky emission is typically subtracted with an accuracy of $\sim$1\%.

Comparing synthetic $g-r$ colours to established photometry, we find a mean offset of 0.043~mag (SAMI data redder) with standard deviation 0.040~mag. The observed $g$-band magnitudes have a mean offset of $-0.047$~mag relative to SDSS photometry (SAMI data brighter), with a scatter of 0.27~mag, although a significant contribution to this scatter comes from difficulties in defining magnitudes for extended, irregular sources. The median full width at half maximum (FWHM) of the final point spread function in the datacubes is 2\farcs1, with a typical range from 1\farcs5 to 3\farcs0.

\subsection{Ancillary data}

The SAMI Galaxy Survey is backed up by a wide range of ancillary datasets. By targetting the GAMA fields, we have available deep spectroscopic data as well as imaging from current and future surveys ranging from the UV to radio, including GALEX MIS, VST KiDS, VISTA VIKING, WISE, Herschel-ATLAS, GMRT and ASKAP \citep{Driver11}. The spectroscopic information is crucial for robustly defining the environments of the SAMI Galaxy Survey targets \citep{Brough13}, while the multi-wavelength imaging data provides a wealth of information on processes (e.g.\ dust-obscured star formation, radio AGN) that may not be apparent in the optical data. The cluster sample also has a variety of ancillary data including archival X-ray observations from {\it XMM-Newton} and {\it Chandra}, and we have carried out a dedicated redshift survey to characterise the clusters and give robust classifications of cluster membership (Owers et al., in prep.).


\section{Early Data Release}

\label{sec:edr}

We have recently made available a set of fully calibrated datacubes for 107 galaxies, forming the SAMI Galaxy Survey Early Data Release (EDR; \citealt{Allen14b}). The galaxies were selected from the GAMA regions of the survey. Also available are the datacubes of the corresponding secondary standard stars and a table of ancillary data including quantities such as stellar mass, effective radius and surface brightness. All data can be downloaded from the SAMI Galaxy Survey EDR website\footnote{\tt http://sami-survey.org/edr}.

\section{Conclusions}

The SAMI Galaxy Survey now includes $\sim$1000 galaxies with IFS observations, and will grow to $\sim$3400 galaxies by mid-2016. This unique dataset is providing new insight into a broad range of topics in galaxy evolution, from star formation in dwarf galaxies to stellar kinematics in galaxy clusters. As the sample grows, further results will come into reach based on robust statistical analyses of the galaxy population, something that has rarely been possible with IFS data in the past. As a starting point, we have publicly released the datacubes for 107 galaxies, allowing the research community access for scientific investigations and preparation for the full dataset.

\section*{Acknowledgements}

\noindent
JTA acknowledges the award of an ARC Super Science Fellowship and a SIEF John Stocker Fellowship.


\begin{thebibliography}{99}

\bibitem[\protect\citeauthoryear{{Allen} et~al.,}{{Allen}
  et~al.}{2014a}]{Allen14a}
{Allen} J.~T.,  et~al., 2014a, Astrophysics Source Code Library, ascl:1407.006

\bibitem[\protect\citeauthoryear{{Allen} et~al.,}{{Allen}
  et~al.}{2014b}]{Allen14b}
{Allen} J.~T., et~al., 2014b, \mnras\ submitted, arXiv:1407.6068

\bibitem[\protect\citeauthoryear{{Brough} et~al.,}{{Brough}
  et~al.}{2013}]{Brough13}
{Brough} S.,  et~al., 2013, \mnras, 435, 2903

\bibitem[\protect\citeauthoryear{{Bryant}, {Bland-Hawthorn}, {Fogarty},
  {Lawrence} \& {Croom}}{{Bryant} et~al.}{2014a}]{Bryant14a}
{Bryant} J.~J.,  {Bland-Hawthorn} J.,  {Fogarty} L.~M.~R.,  {Lawrence} J.~S.,
   {Croom} S.~M.,  2014a, \mnras, 438, 869

\bibitem[\protect\citeauthoryear{{Bryant} et~al.,}{{Bryant}
  et~al.}{2014b}]{Bryant14b}
{Bryant} J.~J.,  et~al., 2014b, \mnras\ submitted, arXiv:1407.7335

\bibitem[\protect\citeauthoryear{{Cappellari} et~al.,}{{Cappellari}
  et~al.}{2011}]{Cappellari11b}
{Cappellari} M.,  et~al., 2011, \mnras, 416, 1680

\bibitem[\protect\citeauthoryear{{Croom} et~al.,}{{Croom}
  et~al.}{2012}]{Croom12}
{Croom} S.~M.,  et~al., 2012, \mnras, 421, 872

\bibitem[\protect\citeauthoryear{{Driver} et~al.,}{{Driver}
  et~al.}{2009}]{Driver09}
{Driver} S.~P.,  et~al., 2009, Astronomy and Geophysics, 50, 12

\bibitem[\protect\citeauthoryear{{Driver} et~al.,}{{Driver}
  et~al.}{2011}]{Driver11}
{Driver} S.~P.,  et~al., 2011, \mnras, 413, 971

\bibitem[\protect\citeauthoryear{{Fogarty} et~al.,}{{Fogarty}
  et~al.}{2012}]{Fogarty12}
{Fogarty} L.~M.~R.,  et~al., 2012, \apj, 761, 169

\bibitem[\protect\citeauthoryear{{Fogarty} et~al.,}{{Fogarty}
  et~al.}{2014}]{Fogarty14}
{Fogarty} L.~M.~R.,  et~al., 2014, \mnras, 443, 485

\bibitem[\protect\citeauthoryear{{Ho} et~al.,}{{Ho}  et~al.}{2014}]{Ho14}
{Ho} I.-T.,  et~al., 2014, \mnras\ accepted, arXiv:1407.2411

\bibitem[\protect\citeauthoryear{{S{\'a}nchez} et~al.,}{{S{\'a}nchez}
  et~al.}{2012}]{Sanchez12}
{S{\'a}nchez} S.~F.,  et~al., 2012, \aap, 538, A8

\bibitem[\protect\citeauthoryear{{Sharp} et~al.,}{{Sharp}
  et~al.}{2006}]{Sharp06}
{Sharp} R.,  et~al., 2006, in SPIE Conference Series Vol.~6269

\bibitem[\protect\citeauthoryear{{Sharp} et~al.,}{{Sharp}
  et~al.}{2014}]{Sharp14}
{Sharp} R.,  et~al., 2014, \mnras\ submitted, arXiv:1407.5237

\bibitem[\protect\citeauthoryear{{York} et~al.,}{{York}  et~al.}{2000}]{York00}
{York} D.~G.,  et~al., 2000, \aj, 120, 1579





\end{thebibliography}
\end{document}